\documentclass[10pt]{article}

 \usepackage{amsmath,amsthm,amssymb}
 \usepackage{makeidx,epsfig,lscape}
 \usepackage{color,colortbl}
 \usepackage{fancyhdr}
 \usepackage{xcolor,pict2e}
 \usepackage{multirow}
 \pdfoutput=1
 \renewcommand{\headrulewidth}{0pt}
 \renewcommand{\footrulewidth}{0.5pt}

 \definecolor{myaqua}{rgb}{0.0,0.5,0.55}
 \definecolor{lightaqua}{rgb}{0.75,0.95,0.95}

 \usepackage[colorlinks = true,
            linkcolor = myaqua,
            urlcolor  = blue,
            citecolor = myaqua]{hyperref}

\usepackage{caption}
\usepackage{floatrow}
\def\lin#1#2{\textcolor[rgb]{0.6,0.6,0.6}{\vspace*{#1mm} \hrule
   height 3 pt \vspace*{#2mm}}}
\def\bt{\begin{tabular}}
\def\et{\end{tabular}}
\def\and{\mbox{ and }}

\def\1{{\bf 1}}

 \def\sectionn#1{\refstepcounter{section}{\color{myaqua}

 \vskip 6mm

 \noindent\Large\bf\thesection. #1}

 \vskip 3mm}

 \def\boxx#1#2#3#4#5{
 {\linethickness{#4pt}\put(#1,#5){\color{myaqua}{\line(1,0){#3}}}}
 \multiput(#1,#2)(0,#4){2}{\line(1,0){#3}}
 \multiput(#1,#2)(#3,0){2}{\line(0,1){#4}}
  }

\begin{document}

 \fancyhead[L]{\hspace*{-13mm}
 \bt{l}{\bf Open Journal of *****, 2020, *,**}\\
 Published Online **** 2020 in SciRes.
 \href{http://www.scirp.org/journal/*****}{\color{blue}{\underline{\smash{http://www.scirp.org/journal/****}}}} \\
 \href{http://dx.doi.org/10.4236/****.2020.*****}{\color{blue}{\underline{\smash{http://dx.doi.org/10.4236/****.2014.*****}}}} \\
 \et}
 \fancyhead[R]{\includegraphics{pic1.ps}}

 $\mbox{ }$

 \vskip 12mm

{ 

{\noindent{\huge\bf\color{myaqua}
  Feinberg-Horodecki exact momentum states of\\[2mm] improved deformed exponential-type potential}}
%
\\[6mm]
{\large\bf Mahmoud Farout$^{1,a}$, Ahmed Bassalat$^{1,b}$, Sameer M. Ikhdair$^{1,2}$}}
\\[2mm]
{ 
$^1$ Department of Physics, An-Najah National University, Nablus, Palestine\\
Email: \href{mailto:m.qaroot@najah.edu}{\color{blue}{\underline{\smash{m.qaroot@najah.edu}}}}\\[1mm]
\href{mailto:ahmad.bassalat@najah.edu}{\color{blue}{\underline{\smash{ahmad.bassalat@najah.edu}}}}\\[1mm]
$^2$ Department of Electrical Engineering, Near East University, Nicosia, Northern Cyprus, Mersin 10, Turkey\\
Email:
\href{mailto:sameer.ikhdair@najah.edu}{\color{blue}{\underline{\smash{sameer.ikhdair@najah.edu}}}}
\\[1mm]
$^b$Corresponding author 1, $^{a}$ Corresponding author 2.\\[4mm]  
Received July, 9, 2020
\\[4mm]
Copyright \copyright \ 2020 by author(s) and Scientific Research Publishing Inc. \\
This work is licensed under the Creative Commons Attribution International License (CC BY). \\
\href{http://creativecommons.org/licenses/by/4.0/}{\color{blue}{\underline{\smash{http://creativecommons.org/licenses/by/4.0/}}}}\\
\includegraphics{pic2.ps}

\lin{5}{7}

 { 
 {\noindent{\large\bf\color{myaqua} Abstract}{\bf \\[3mm]
 \textup{We obtain the quantized momentum eigenvalues, $P_n$, and the momentum eigenstates for the space-like Schrodinger equation, the Feinberg-Horodecki equation, with the improved deformed exponential-type potential which is constructed by temporal counterpart of the spatial form of these potentials.  We also plot the variations of the improved deformed exponential-type potential with its momentum eigenvalues for few quantized states against the screening parameter. 
 }}}
 \\[4mm]
 {\noindent{\large\bf\color{myaqua} Keywords}{\bf \\[3mm]
 Quantized momentum states; Feinberg-Horodecki equation; the time-dependent improved deformed exponential-type potential. }

 \fancyfoot[L]{{\noindent{\color{myaqua}{\bf How to cite this
 paper:}} M. Farout and S. M. Ikhdair (2020)
Momentum eigensolutions of Feinberg-Horodecki equation with time-dependent screened Kratzer-Hellmann potential.
 ***********,*,***-***}}

\lin{3}{1}

\sectionn{Introduction}

{ \fontfamily{times}\selectfont
 \noindent In studying any physical problem in quantum mechanics we seek to find the solution of the resulting second-order differential equation. The time-dependent Schrödinger equation represents an example that describes quantum-mechanical phenomena, in which it dictates the dynamics of a quantum system. Solving this differential equation by means of any method results in the eigenvalues and eigenfunctions of that Schrödinger quantum system. However, the solution of the time-dependent
 Schrödinger equation analytically is exact and limited to certain problems of spatial coordinate problems  \cite{Park02, Vorobeichik98, Shen03, Feng01}.
 The Feinberg-Horodecki (FH) equation is an equivalent time-momentum equation to the energy-spatial coordinate Schrodinger equation which was derived by Horodecki \cite{Horodecki88} from the relativistic Feinberg equation \cite{Feinberg67}. This equation has been demonstrated in a possibility of describing biological systems \cite{Molski06, Molski10} in terms of the time-like supersymmetric quantum mechanics \cite{Witten81}. The spatial-like solution of the FH equation can be employed to test its relevance
 in different areas of science including physics, biology and medicine \cite{Molski06, Molski10}. Molski obtained the spatial-like states of the time-dependent Morse oscillator potential model in the framework of the FH equation for minimizing the time-energy uncertainty relation and showed that the results are useful for interpreting the formation of the
 specific growth patterns during crystallization process and biological growth \cite{Molski06}. Furthermore, he solved the FH equation with anharmonic oscillators and obtained the space-like quantum supesymmetry for the sake of describing biological systems \cite{Molski10}.
 
 Recently, Bera and Sil found the exact solutions of the FH equation for the time-dependentWei-Hua oscillator and Manning-Rosen potentials by the Nikiforov-Uvarov (NU) method \cite{Bera13}. A simple form of a potential model \cite{Deng57} named Deng-Fan oscillator potential was introduced in 1957. This potential is taking a general Morse potential mesa1998generalized has been studied for it's energy spectrum and wave functions by Refs \cite{Deng57, Wang12, Rong03, Mesa98} and related to the Manning-Rosen potential \cite{Infeld51, Manning33} which is also called
 Eckart potential by some authors \cite{Dabrowska88, Cooper95, Zhang05} or anharmonic potential. This system is well-defined at boundaries where t=0 and t=1. The spatial-like Deng-Fan model is quantitatively very similar to Morse model with correct asymptotic behaviour when inter nuclear separation distance comes to zero \cite{Deng57} and correctly describe the spectrum of diatomic molecules and electromagnetic transition \cite{Dong2008, Hamzavi13a, Oluwadare12}. The FH equation is solved with the time-dependent Deng-Fan oscillator potential model to obtain the exact momentum states by means of the parametric NU method \cite{Hamzavi13b}. 
 
 Recently, Altug and Sever have studied the FH equation with time-dependent Poschl-Teller potential and
 found its space-like coherent states \cite{Arda17}. We also studied the solutions of FH equation for time-dependent mass (TDM) harmonic oscillator quantum system. An appropriate interaction to time-dependent mass is chosen to obtain the correct spectrum of stationary energy. The related spectrum of Harmonic oscillator potential acting on the TDM stationary state energies are found \cite{Eshghi16}. The exact solutions of FH equation under time-dependent Tietz-Wei di-atomic molecular potential have been obtained. In particular, the quantized momentum eigenvalues and
 corresponding wave functions are found in framework of supersymmetric quantum mechanics \cite{Ojonubah16}. The spectra of general molecular potential (GMP) are obtained using asymptotic iteration method within the framework of
 non-relativistic quantum mechanics. The vibrational partition function is calculated in closed form and used to
 obtain thermodynamic functions \cite{Ikot18}.
 
 Recently, we solved the FH equation with the time-dependent Kratzer plus screened Coulomb potential \cite{Farout20a}, we solved FH equation with the time-dependent screened Kratzer-Hellmann potential model \cite{Farout20b}, and very recently we a general time-dependent potential \cite{Farout20c} . In each case, we obtained the approximated eigensolutions of momentum states and wave functions by means of the NU method.
 
 The motivation of this work is to apply the NU method \cite{Nikiforov88} for the general molecular potential having a certain time-dependence. The momentum eigenvalues, $P_n$, of the FH equation and the space-like coherent eigenvectors are obtained. The rest of this work is organized as follows: the NU method is briefly introduced in Section 2. The exact solution of the FH equation for the time-dependent general molecular potential is solved to obtain its quantized momentum states and eigenfunctions in Section 3. 
 We generate the solutions of a few special potentials mainly found from our general form solution in section 4. Finally we present our discussions and conclusions.

\renewcommand{\headrulewidth}{0.5pt}
\renewcommand{\footrulewidth}{0pt}

 \pagestyle{fancy}
 \fancyfoot{}
 \fancyhead{} 
 \fancyhf{}
 \fancyhead[RO]{\leavevmode \put(-90,0){\color{myaqua}M. Farout, S. M. Ikhdair} \boxx{22}{-10}{10}{50}{15} }
 \fancyhead[LE]{\leavevmode \put(0,0){\color{myaqua}M. Farout, S. M. Ikhdair}  \boxx{-45}{-10}{10}{50}{15} }
 \fancyfoot[C]{\leavevmode
 \put(0,0){\color{lightaqua}\circle*{34}}
 \put(0,0){\color{myaqua}\circle{34}}
 \put(-2.5,-3){\color{myaqua}\thepage}}

 \renewcommand{\headrule}{\hbox to\headwidth{\color{myaqua}\leaders\hrule height \headrulewidth\hfill}}

}

\sectionn{Exact Solutions of the FH Equation for the Time-Dependent Improved Deformed Exponential-type Potential}

{ \fontfamily{times}\selectfont
 \noindent
The Nikiforov-Uvarov (NU) method (see appendix \ref{Appendix}) will be used to find the exact solutions of FH equation for the improved deformed exponential-type (IDEP) which results in momentum eigenvalues and their eigenstates.

The time-dependent of the improved deformed exponential-type potential is given by \cite{Okorie20}

\begin{equation}
V(t)=D_e\left[1-\frac{q-e^{2\alpha(t_e-t_0)}}{q-e^{2\alpha(t-t_0)}}\right]^2.
\end{equation}

where $t_0$ and $\alpha$  are adjustable real potential parameters. $q$ is a dimensionless parameter, $D_e$ is the dissociation energy and $t_e$ the equilibrium time point. The IDEP is reduces to the improved Tietz potential if $2\alpha$ is replaced by $\alpha$ and $-q e^{2\alpha t_0}$ by $h$.
If the IDEP potential is substituted in FH equation, one obtains
\begin{equation}
\left[-\frac{\hbar^2}{2mc^2}\frac{d^2}{dt^2}+ D_e\left(1-\frac{q-e^{2\alpha(t_e-t_0)}}{q-e^{2\alpha(t-t_0)}}\right)^2\right]\psi_n(t) = cP_n \psi_n(t).
\end{equation}
Now, let $s=\frac{1}{q}e^{2\alpha(t-t_0)}$, where s $\epsilon$($\frac{1}{q}e^{-2\alpha t_0}$, $\infty$), we get
\begin{equation}
\frac{d^2\psi_n(s)}{ds^2} + \frac{(1-s)}{s(1-s)}\frac{d\psi_n(s)}{ds}+\frac{-\zeta_1^2-\zeta_3s+\zeta_2s^2}{s^2(1-s)^2}\psi_n(s)=0,
\label{eq:differential eq to solve}
\end{equation}
where
\begin{equation}
\beta=-\frac{2\hbar\alpha^2}{mc^2},
\label{eq:beta}
\end{equation}
\begin{equation}
L=\frac{D_e}{\beta},
\end{equation}
\begin{equation}
M=-\frac{cP_n}{\beta},
\end{equation}
\begin{equation}
A=\frac{D_e e^{4\alpha(t_e-t_0)}}{q^2\beta},
\label{eq:A}
\end{equation}

\begin{equation}
C=\frac{2D_e e^{2\alpha(t_e-t_0)}}{q \beta},
\label{eq:c}
\end{equation}

\begin{equation}
\zeta_1^2=-(A+M),
\label{eq:zeta1}
\end{equation}
\begin{equation}
\zeta_2=-(C+2M),
\end{equation}
and
\begin{equation}
\zeta_3=-(L+M).
\end{equation}
After comparing equation (\ref{eq:differential eq to solve}) with equation (\ref{eq: NU-equ}), one obtains \\
$\tilde{\tau}(s)=1-s$, $\sigma(s)= s(1-s)$, and $\tilde{\sigma}(s)=-\zeta_1^2+\zeta_2s+-\zeta_3 s^2$.\\
When these values are substituted in equation $\Pi(s)= \frac{\sigma^{'}-\tilde{\tau}}{2}\pm \sqrt{(\frac{\sigma^{'}-\tilde{\tau}}{2})^2-\tilde{\sigma}+ k\sigma}$ (see NU method \cite{Farout20a, Farout20b, Farout20c}), we get
\begin{equation}
\Pi(s)= -\frac{s}{2}\pm \sqrt{\left(\frac{1}{4}+\zeta_3-k\right)s^2+ (k-\zeta_2)s+\zeta_1^2}.
\label{eq: pi solving}
\end{equation}
As mentioned in the NU method, the discriminant under the square root, in equation (\ref{eq: pi solving}), has to be zero, so that the expression of $\Pi (s)$ becomes the square root of a polynomial of the first degree. This condition can be written as
\begin{equation}
\left(\frac{1}{4}+\zeta_3-k\right)s^2+ (k-\zeta_2)s+\zeta_1^2=0.
\end{equation}
After solving this equation, we get
\begin{equation}
s= \frac{-(k+\zeta_3)\pm\sqrt{(k+\zeta_3)^2-4\zeta_1^2(\frac{1}{4}-\zeta_2-k)}}{2(\frac{1}{4}-\zeta_2-k)}.
\label{eq: s}
\end{equation}
Then, for our purpose we assume that
\begin{equation}
(k+\zeta_3)^2-4\zeta_1^2\left(\frac{1}{4}-\zeta_2-k\right)=0.
\end{equation}
Arranging this equation and solving it to get an expression for k which is given by the following,
\begin{equation}
k_\pm = \zeta_2-2\zeta_1^2\pm 2\zeta_1\left(\frac{1}{R}-\frac{1}{2}\right),
\end{equation}
where the expression between the parentheses is given by
\begin{equation}
\frac{1}{R}-\frac{1}{2}= \sqrt{C-A-L+\frac{1}{4}}.
\label{eq:1/R-1/2}
\end{equation}
where the parameters in this equation must be selected to let R be real and the results have physical meanings. 
If we substitute $k_-$ into equation (\ref{eq: pi solving}) we get a possible expression for $\Pi(s)$, which is given by 
\begin{equation}
\Pi(s) =\zeta_1 -(\zeta_1+\frac{1}{R})s,
\label{eq: pi(s) solution}
\end{equation}

this solution satisfy the condition that the derivative of $\tau(s)$ is negative. Therefore, the expression of $\tau(s)$ which satisfies these conditions can be written as
\begin{equation}
\tau(s) =1-s +2\zeta_1-2s(\zeta_1+\frac{1}{R}).
\label{eq: tau(s) result}
\end{equation}
Now, substituting the values of $\tau^{'}_-(s)$, $\sigma^{''}(s)$, $\Pi^{'}_-(s)$ and $k_-$ into equations (\ref{eq: lambda1}) and (\ref{eq: lambda2}), we obtain
\begin{equation}
\lambda_n=\zeta_2-2\zeta_1^2-2\zeta_1(\frac{1}{R}-\frac{1}{2})-(\frac{2 \zeta_1}{R}+\zeta_1),
\label{eq:lambda_n solution}
\end{equation}
and 
\begin{equation}
\lambda= \lambda_n= n(n-1)+2n(1+ \frac{2}{R})+ 2n \zeta_1.
\label{eq:lambda-solution}
\end{equation}
Now, from equations (\ref{eq:lambda_n solution}) and (\ref{eq:lambda-solution}), we get the eigenvalues of the quantized momentum as
\begin{equation}
P_n= \frac{1}{c}\left(A+\beta\left[\frac{2A-C-n(n+1)-\frac{2n+1}{R}}{2(n+\frac{1}{R})}\right]^2\right)
\label{eq:eigenvalues of Pn}
\end{equation}
where $\beta$, A and C are defined in equations \ref{eq:beta}, \ref{eq:A} and \ref{eq:c} respectively.

Due to the NU method used in getting the eigenvalues, the polynomial solutions of the hypergeometric function $y_n(s)$ depend on the weight function $\rho(s)$ which can be determined by solving $\sigma (s) \rho^{'}(s) +[\sigma(s) - \tau(s)] \rho(s) =0$ to get 
\begin{equation}
\rho(s)= s^{2\zeta_1}(1-s)^{(\frac{2}{R})-1}.
\label{eq:rho-result}
\end{equation}
Substituting $\rho(s)$ into $y_n(s)= \frac{A_n}{\rho(s)} \frac{d^n}{ds^n} [\sigma^n(s) \rho (s)]$, we get an expression for the wave functions as
\begin{equation}
y_n(s)= A_n s^{-2\zeta_1}(1-s)^{-(\frac{2}{R}-1)}\frac{d^n}{ds^n}\left[s^{n+2\zeta_1}(1-s)^{n+\frac{2}{R}-1}\right],
\label{eq: yn 2}
\end{equation}
where $A_n$ is the normalization constant. Solving equation (\ref{eq: yn 2}) gives the final form of the wave function in terms of the Jacobi polynomial $P_n^{(\alpha, \beta)}$ as follows,
\begin{equation}
y_n(s)= A_n n! P_n^{(2\zeta_1, \frac{2}{R}-1)}(1-2s).
\label{eq: yn last solution}
\end{equation}
Now, substituting $\Pi_-(s)$ and $\sigma(s)$ into $\sigma(s)= \Pi(s) \frac{\phi_n(s)}{{\phi_n}^{'}(s)}$ then solving it we obtain
\begin{equation}
\phi_n(s)= s^{\zeta_1}(1-s)^\frac{1}{R}.
\label{eq: phi solutin}
\end{equation}
Substituting equations (\ref{eq: yn last solution}) and (\ref{eq: phi solutin})in $\psi_n(s)= \phi_n(s) y_n(s)$, one obtains,
\begin{equation}
\psi_n(s)= B_n s^{\zeta_1}(1-s)^\frac{1}{R} P_n^{(2\zeta_1, (\frac{2}{R})-1)}(1-2s),
\end{equation}
where $B_n$ is the normalization constant. 
}

\sectionn{Numerical results and discussion}

{ \fontfamily{times}\selectfont
 \noindent
We compute the momentum eigenvalues of
time dependent improved deformed exponential-type potential for some diatomic molecules like $CO$, $N_2$, $H_2$ and $LiH$. This was done using the spectroscopic parameters displayed in Table \ref{table:paraeter}.

\begin {table}[H]
{\scriptsize
	\caption {Spectroscopic parameters of the various diatomic molecules \cite{Okorie20}} \label{tab:Parameters} 
	\vspace*{12pt}
	\begin{center}
		\begin{tabular}{c  c   c  c  c  c}
			
			\hline
			\\
			Molecule & $D_e$ (eV) & $t_e (ns)$  & $\mu$ (a.m.u)&$t_0 (ns)$&$q$ \\\\
			\hline\\
			CO & 10.84514471 &  1.1283 & 6.860586000 & 1.128300118 & -0.6544806294  \\\\
			
			N$_2$ & 9.9051 &  1.0970 & 7.0034& 1.097000113 &-0.3543700921   \\\\
			
			H$_2$  & 4.7446 &  0.7416 & 0.5039& 0.7416001485 &-0.3236073943   \\\\
			
			LiH & 2.5155 &  1.5955 & 0.8801& 1.595500403 & -0.3326882575  \\
			
			\end {tabular}
			\label{table:paraeter}
		\end{center}
		\vspace*{-2pt}}
\end{table}
\begin{figure}[H]
	\includegraphics[width=1.05\linewidth]{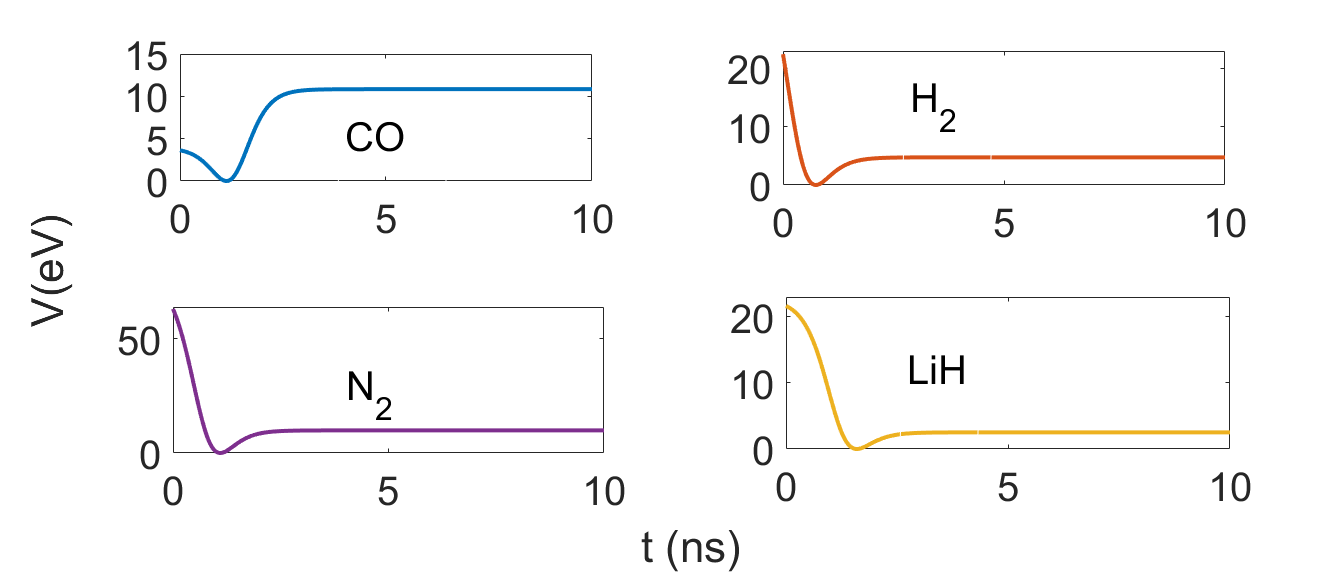}
	\caption[IDEP Potential]{Improved deformed exponential-type potential (IDEP) for diatomic molecules. The parameters used are presented in Table \ref{table:paraeter}, and $\alpha =0.5~ (ns)^{-1}$}
	\label{fig:IDEP_time}
\end{figure}

\begin{figure}[H]
	\includegraphics[width=1.05\linewidth]{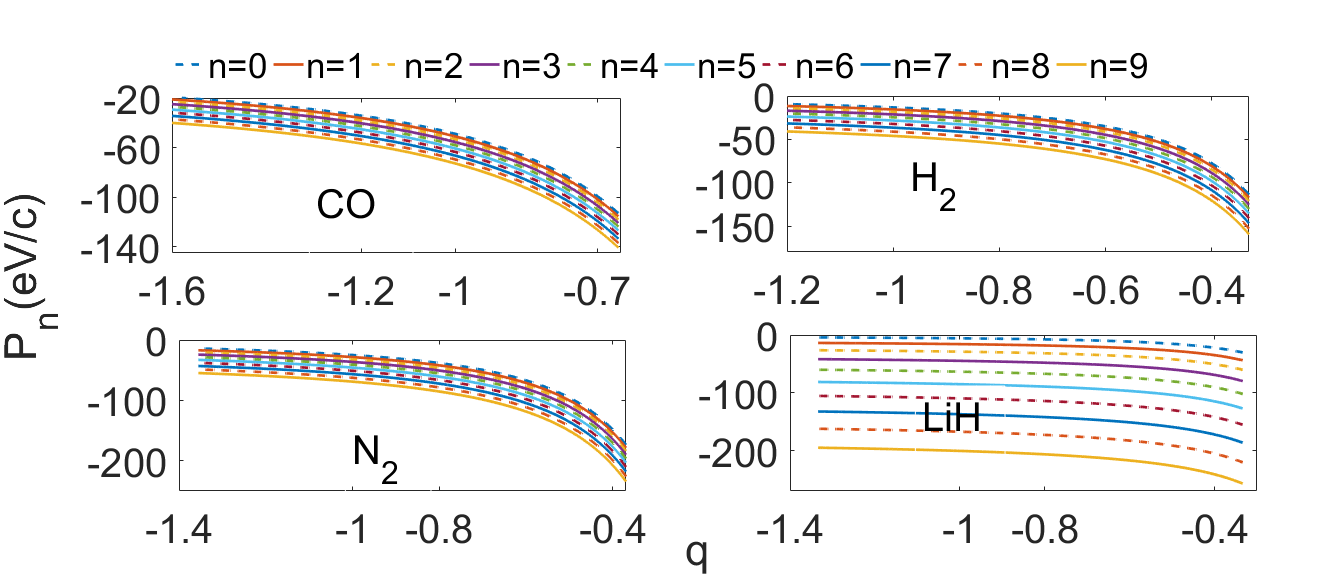}
	\caption[IDEP Potential]{FH quantized momentum eigenvalues for the improved deformed exponential-type potential plotted vs $q$ for diatomic molecules.The parameters used are presented in Table \ref{table:paraeter}, and $\alpha =0.5~ (ns)^{-1}$.}
	\label{fig:IDEP_Pn_q}
\end{figure}

\noindent Figure \ref{fig:IDEP_time} shows the variation of the time-dependent improved deformed exponential-type potential (IDEP) well for four different diatomic molecules at small times. Hence, this  potential well changes from 20 $eV$ to nearly 5 $eV$ for the diatomic molecules $H_2$ and $LiH$ whereas it changes from 50 $eV$ to 5 $eV$ for $N_2$ diatomic molecule. However, $CO$ diatomic molecule has a unique behavior, it varies from 5 $eV$ to 10 $eV$. 
In Fig. \ref{fig:IDEP_Pn_q}, we examined the variation of the quantized momentum states $P_n$ of IDEP against the screening parameter $q$ for various diatomic molecules. It is seen that the momentum of the present potential model decreases monotonically from zero, for $H_2$, $N_2$ and $LiH$ whereas for $CO$ it shows different behavior as it decreases from -20 $eV/c$, against the screening parameter $q$ for various values of states, $n$.
Figure \ref{fig:IDEP_Pn_alpha} shows the variation of $P_n$ in the field of IDEP against the exponential parameter $\alpha$ for various diatomic molecules. The diatomic molecules exhibit different features; for various values of n. It is obvious from figure that $P_n$ increases monotonically with increasing $\alpha$ for $CO$ and $N_2$ when particle is subjected to the aforementioned system. The reverse case happens with $H_2$ and $LiH$ diatomic molecules where $P_n$ decreases monotonically from a value close to zero with increasing $\alpha$.

\begin{figure}[H]
	\includegraphics[width=1.05\linewidth]{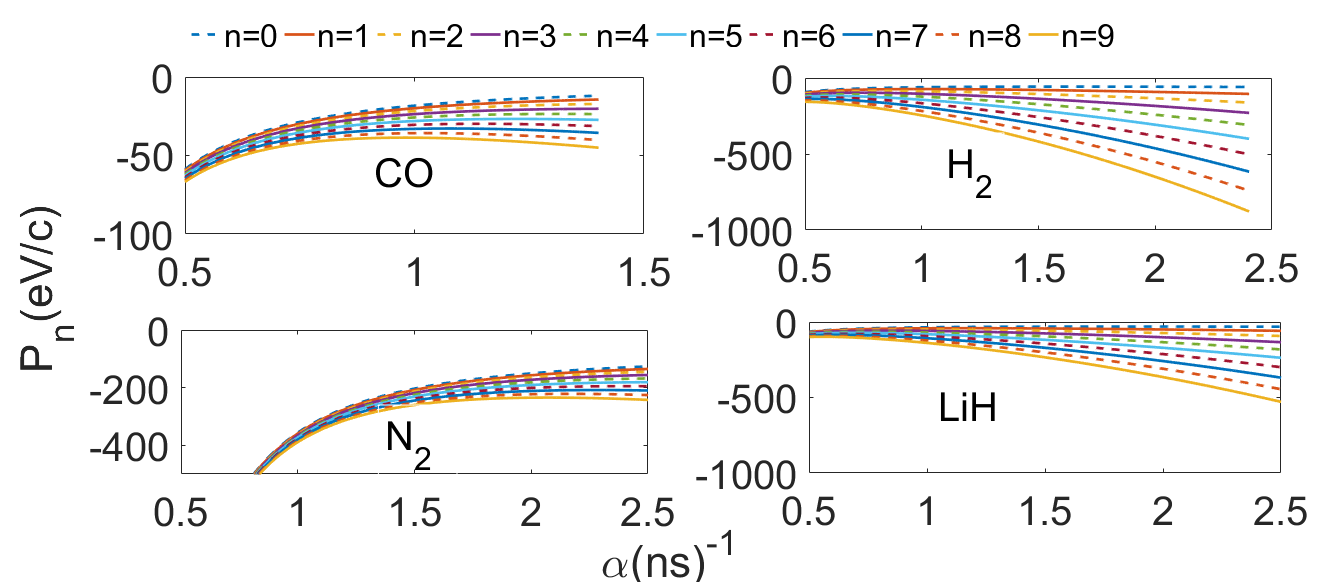}
	\caption[IDEP Potential]{FH quantized momentum eigenvalues for the improved deformed exponential-type potential plotted vs $\alpha$ for diatomic molecules. The parameters used are presented in Table \ref{table:paraeter}.}
	\label{fig:IDEP_Pn_alpha}
\end{figure}
\begin{figure}[H]
	\includegraphics[width=1.05\linewidth]{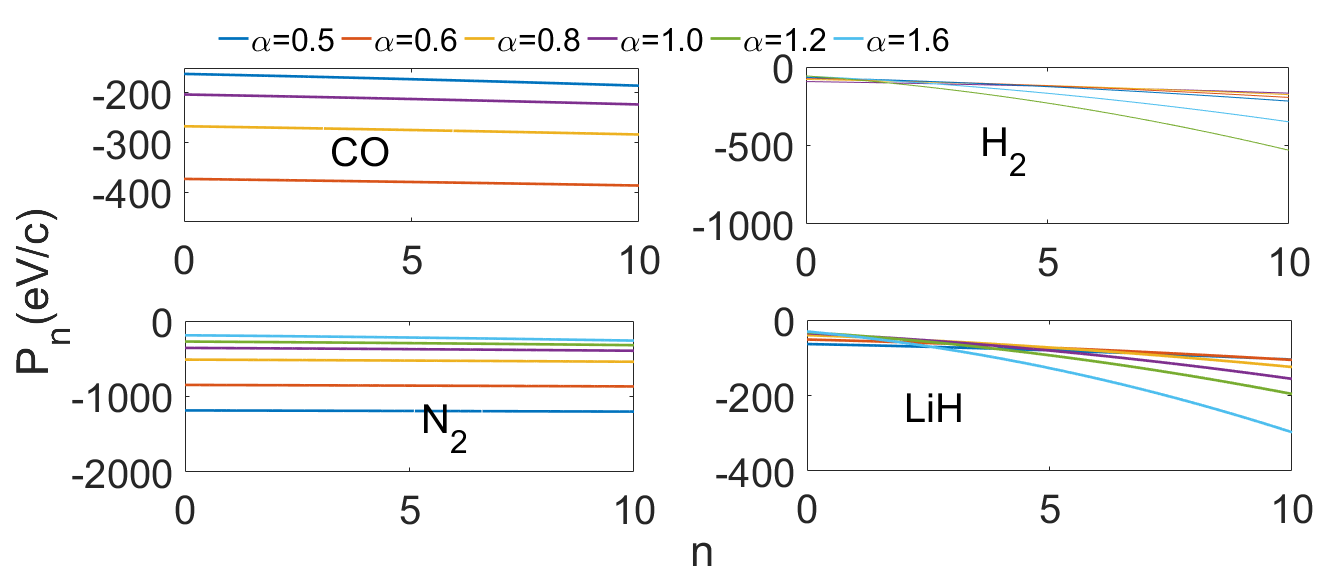}
	\caption[IDEP Potential]{FH quantized momentum eigenvalues for the improved deformed exponential-type potential plotted vs $n$ for diatomic molecules. The parameters used are presented in Table \ref{table:paraeter}.}
	\label{fig:IDEP_Pn_n}
\end{figure}

Finally, Fig. \ref{fig:IDEP_Pn_n} shows the behavior of $P_n$ against state $n$ for various diatomic molecules subjected to the field of IDEP with various values of $\alpha$. It is seen that $P_n$ for $H_2$ and $LiH$ decreases monotonically from zero with increasing $n$. However, in the case of $CO$ and $N_2$ diatomic molecules, it is seen that $P_n$ decreases slightly linearly with increasing $n$.
}
\section{Conclusions}
{ \fontfamily{times}\selectfont
	\noindent

We solved the Feinberg-Horodecki (FH) equation for the time-dependent improved deformed exponential-type potential via Nikiforov-Uvarov (NU) method. We got the exact quantized momentum eigenvalues solution of the FH equation. It is therefore, worth mentioning that the method is elegant and powerful. Our results can be applied in biophysics and other branches of physics.
We find that our analytical results are in good agreement with other findings in literature. We have shown the behaviors of the improved deformed exponential-type potential  against screening parameters.
Further, taking spectroscopic values for the potential parameters, we plotted the quantized momentum of few states against the screening parameter for diatomic molecules. Our results are good agreements with the energy bound states. 
}

\appendix

\section*{Appendix: Methodology}
\label{Appendix}
{ \fontfamily{times}\selectfont
 \noindent

The Nikiforov-Uvarov (NU) \cite{Nikiforov88} method is an efficient technique usually employed to reduce the second-order differential equation into a general form of a hypergeometric type equation. Therefore, any second order differential equation can be transformed, via an appropriate coordinate transformation $s=s(t)$, into a standard form:
\begin{equation}
\psi_n^{''}(s)+ \frac{\tilde{\tau}(s)}{\sigma(s)} \psi_n^{'}(s)+ \frac{\tilde{\sigma}(s)}{\sigma^2(s)} \psi_n(s)=0,
\label{eq: NU-equ}
\end{equation}
where $\sigma (s)$ and $\tilde{\sigma}(s)$ are polynomials, at most second-order, and $\tilde{\tau}(s)$ is of a first-order polynomial. 
To follow the method in details the reader is advised to follow \cite{Nikiforov88,Farout20a,Farout20b, Farout20c}.
The eigenvalues equation can be found simply by solving the equations (\ref{eq: lambda1}) and (\ref{eq: lambda2}). Where
\begin{equation}
\lambda = \lambda_n = -n \tau^{'}(s) - \frac{n(n-1)}{2} \sigma^{''}(s),
\label{eq: lambda1}
\end{equation}
and
\begin{equation}
\lambda = \lambda_n= k+ \Pi^{'}(s),
\label{eq: lambda2}
\end{equation}
with n=0, 1, 2, ......, and $\tau(s)$ is a polynomial with a negative first derivative to generate an appropriate solution for the hypergeometric equation. And
Further, $\Pi(s)$ is a polynomial which depends on the transformation function s(t) and k should be determined to calculate $\Pi(s)$, for which the discriminant under the square root  is set to zero, in order to let $\Pi(s)$ to be a first order polynomial. 
}

\vskip 3mm

 \noindent\Large\bf Acknowledgments}

 \vskip 3mm

{ \fontfamily{times}\selectfont
 \noindent
 We thank the Editor and the referees for their valuable comments.
 This research is funded by Winter School in High Energy Physics in Palestine (WISHEPP). This generous support is greatly appreciated.

 {\color{myaqua}

}}
\end{document}